\documentclass[twocolumn,prb,showpacs]{revtex4}
\usepackage{graphicx}
\usepackage{epsfig}
\usepackage{amsmath}
\usepackage{amssymb}
\usepackage{gensymb}
\usepackage{bm}
\usepackage[usenames]{color}

\newcommand{\beq}{\begin{eqnarray}}
\newcommand{\eeq}{\end{eqnarray}}

\begin{document}

\title{Mechanism of Ferroelectricity in $d^3$ Perovskites - a Model Study}

\author{Paolo Barone$^{1}$}
%\affiliation{CNR-SPIN 67100 L'Aquila, Italy}
\author{Sudipta Kanungo$^{2}$}
%\affiliation{S. N. Bose National Centre for Basic Sciences, Sector III, Block
%JD, Salt Lake, Kolkata 700 098, India}
\author{Silvia Picozzi$^{1}$}
\author{Tanusri Saha-Dasgupta$^{2}$}
%\affiliation{S. N. Bose National Centre for Basic Sciences, Sector III, Block
%JD, Salt Lake, Kolkata 700 098, India}

\affiliation{1. CNR-SPIN 67100 L'Aquila, Italy}
\affiliation{2. S. N. Bose National Centre for Basic Sciences, Sector III, Block
JD, Salt Lake, Kolkata 700 098, India}

\begin{abstract}
By means of a model Hamiltonian approach we study the role of volume expansion, Hund's coupling and electron correlation
in the standard hybridization mechanism for ferroelectricity in cubic CaMnO$_3$, a prototypical non-$d^0$ perovskite. Our
results establish that the ferroelectric instability arises from a subtle balance between different energy contributions, explaining 
the origin of its enhancement under negative pressure. Expansion of volume is found to cause a strong reduction of the elastic energy,
while leaving almost unchanged the tendency of Mn states to form covalent bonds with the surrounding oxygens. Hund's coupling with 
local spins of magnetic cations can reduce and even suppress the instability towards the ferroelectric state.
\end{abstract}


\maketitle

\section{Introduction}\label{section:introduction}

Since the revival of interest in the magnetoelectric effect
due to the possibility of multiferrocity\cite{fiebig} with 
simultaneous presence of ferroelectric and magnetic order, many attempts have been made
to create multiferroic materials focusing on the wide variety of transition metal ABO$_3$
perovskites. 

The classical ferroelectrics as well as many magnetic materials belong\cite{mitsui,goodenough}
to this class of compounds. However, overlap between ferroelectric perovskites and magnetic perovskites
are rare, with the exception of compounds like BiFeO$_3$ and BiMnO$_3$. This lead to the suggestion 
of general ``exclusion'' rule for perovskites\cite{hill,filippetti,khomski1,khomski2}, according to which
ferroelectricity occurs for $d^0$ B site transition-metal ($TM$) cation\cite{matthias}, while
the magnetism requires $d^n$ B site cations with $n \ne 0$. This mutual exclusion of ferroelectric and
magnetic properties is rationalized by the cases of conventional ferroelectric perovskites like, e.g.,
BaTiO$_3$, KNbO$_3$ or Pb(ZrTi)O$_3$, for which the driving force behind the ferroelectric distortion 
comes from the tendency of $TM$ empty $d$ states to establish a strong covalency with the 
surrounding oxygens.\cite{khomski1,cohen_nature} The hybridization driven mechanism is maximally 
efficient when the antibonding states, with predominant $TM$ $d$ character, are completely empty, hence 
for $d^0$ B site perovskites. The above mentioned exclusion principle is violated for cases like BiFeO$_3$ and BiMnO$_3$,
for which the polarization arises in A sublattice due to the lone-pair stereochemical activity of Bi 
$6s$ electrons\cite{seshadri} and magnetism arises in B sublattice, with no apparent coupling between the two order
parameters. For technological purpose, on the other hand, it is highly desirable to invoke coupling between the
two order parameters, so that magnetism can be controlled by electric field and polarization by magnetic field.
Along this line of thought, a number of magnetically induced multiferroic perovskites 
have been synthesized and studied, in which the complicated magnetic
structure either with non-collinear spiral spin configuration\cite{kimura,sergienko_prb,knb} or
collinear $\uparrow\,\uparrow\,\downarrow\,\downarrow$ E-type
antiferromagnetic\cite{choi,sergienko_prl,picozzi_prl} structure, breaks the inversion symmetry and induces polarization.
Also non-trivial charge distributions in systems displaying both charge
and/or orbital order  may significantly contribute to the onset of polar ferroelectric
phases\cite{khomskiibrink,barone_prl,barone_prb}, even concomitantly with magnetism.
These compounds, known as improper ferroics, normally do not exhibit off centric movement of magnetic cations.
While such electronically driven mechanism has generated lot of academic interest, the resulting polarization
is often tiny, precluding technological applications. Search is, therefore, on to obtain multiferroic materials
showing magnetoelectric coupling with reasonably high transition temperature and polarization values.

Since the conventional lattice driven mechanism of polarization with off centric movement of B site cations 
arises due to delicate balance between the hybridization energy gain and the repulsive electrostatic energy,
the general applicability of the ``exclusion principle'' for a $d^n$ perovskite with $n < 5$ is difficult to
predict a-priori. $d^n$ perovskite, if can be made to a conventional ferroelectric, is expected
to exhibit magnetoelectric coupling as well as presumably large values of polarization. Mn based perovskites like
CaMnO$_3$, BaMnO$_3$ or SrMnO$_3$ may be considered as possible candidates, since B site cation in this case 
is in a $d^3$ electronic configuration, with electrons occupying, within a cubic crystal field, the 3-fold
degenerate $t_{2g}$ states, leaving the $e_g$ states unoccupied. Noticeably, like
conventional $d^0$ ferroelectric perovskites, they exhibit anomalously large Born effective charges (BECs) of Mn
ions.\cite{ederer,ghosez,rondinelli}
The study by Ederer {\it et al.}\cite{ederer} showed that the largest anomalous contributions to BECs can be understood in terms 
of a charge transfer between $TM$ $d$ and O $p$ states along the $\sigma-$type channel, suggesting that the hybridization mechanism 
is mostly $e_g$-driven. Attempts have been made to induce ferroelectricity in these compounds through increase of volume 
by applying negative or chemical pressure\cite{ghosez,rondinelli}, or in epitaxially strained thin films\cite{leerabe}. Interestingly, 
the polarization in these predicted systems may be larger than that typically expected in magnetically-induced multiferroics such as
TbMnO$_3$ ($P\lesssim 0.1~ \mu$C/cm$^2$) or HoMnO$_3$ ($P\sim 0.5-2~ \mu$C/cm$^2$). Rather large polarization value of $P=54~
\mu$C/cm$^2$ has been predicted for epitaxially strained SrMnO$_3$.\cite{leerabe} 

A microscopic understanding of this interesting behavior demands further study, in particular what makes ferroelectric instability 
stronger at larger lattice constants. Hybridization effect is expected to be weaker for increased $TM$ - O bondlengths and therefore
would disfavor the hybridization-driven ferroelectricity even further. The role of Hund's coupling between   
$t_{2g}$ and $e_g$ spins, which has been suggested\cite{khomski1} to prevent ferroelectricity in compounds like CaMnO$_3$, needs
to be also understood in the context of increased lattice constant. In the present study, we investigated this issue 
in a model Hamiltonian based approach. The model Hamiltonian approach, as opposed to the first-principles density functional theory (DFT) 
calculations, has the advantage of analysing the different energy contributions individually and providing a microscopic 
understanding of the influence of various different factors. However, since the ferroelectric instability arises due to delicate 
balance between different competing energies, we employ DFT tools for accurate estimates of different energies. Specifically, we have
used muffin-tin orbital (MTO) based NMTO-downfolding technique\cite{andersen_nmto} to estimate the hopping interactions between
nearest neighbor Mn and oxygen atoms, and that between next nearest neighbor oxygens. The energy associated with the stiffness
of the off-centric, ferroelectric movement has been estimated using total-energy DFT calculations carried out employing 
the plane-wave pseudopotential method as implemented within VASP.\cite{vasp} We assumed the high temperature cubic crystal 
structure of CaMnO$_3$, which is the prototypical cubic $TM$ perovskite crystal structure. We have not considered in our study
the antiferrodistortive tilt and rotation of MnO$_6$ octahedra which have been found\cite{ghosez,rondinelli} to compete with the 
ferroelectric instability. Our study, therefore, focuses on the competition between the pair-breaking Hund's rule coupling, the
hybridization effect and the short-range, electrostatic repulsion in destabilizing or stabilizing ferroelectricity in a prototypical cubic
perovskite structure, and the effect of volume expansion on it.

\section{Methods and computational details}\label{section:model}

For our study, we used a minimal model designed to reproduce the tendency of empty $e_g$ states to establish covalent bonds with
surrounding oxygens. Neglecting the $t_{2g}$ electrons motion, the model should describe the hopping between filled
oxygen $p$ state and empty Mn two-fold $e_g$ states on the background of localized $t_{2g}$ spins.
Within the cubic crystal field, the $e_g$ states establish prevalently $\sigma$ bonds with O $p$ states 
parallel to the bond direction\cite{footnote}, as confirmed by Wannier function analysis\cite{ederer,satpathy}. We therefore 
considered a 5-band $d-p$ model, given by
$
H=H_d+H_p+H_{dp},
$
where
\beq\label{ham:mn}
H_d=\varepsilon^d\,\sum_{i,\alpha,\sigma}\,\,d^{\dagger}_{i\alpha\sigma}\,
d^{\phantom{\dagger}}_{i\alpha\sigma} \,-\,J_H\,\sum_i\, {\bm s}_i\cdot{\bm S}_i + H_{e-e}
\eeq
takes into account the local physics on $TM$ site,
\beq\label{ham:o}
H_p=\varepsilon^p_0\,\sum_{j,\sigma}p^{\dagger}_{j\sigma}\,p^{\phantom{\dagger}}_{j\sigma} +
\sum_{j,j',\sigma} t^{pp}_{j,j'}\,p^{\dagger}_{j\sigma}\,p^{\phantom{\dagger}}_{j'\sigma}
\eeq
describes the oxygen sites, and
\beq
H_{dp}=\sum_{i,j,\alpha,\sigma}\,V_{i,j}^{\alpha}d^{\dagger}_{i\alpha\sigma}\,p^{\phantom{\dagger}}_{j\sigma}
\eeq
describes the hybridization between Mn and O sites. $d^{\dagger}_{i,\alpha,\sigma}$ creates an electron with spin
$\sigma$ on a Mn at site $i$ and in the orbital state $x^2-y^2$ ($3z^2-r^2$) for $\alpha=1 (2)$, while
$p^\dagger_{j,\sigma}$ creates an electron on O site $j=i\pm \hat{e}/2$, with $\hat{e}$ the
direction unit vector $\hat{x},\hat{y},\hat{z}$, in the orbital state $x,y,z$ respectively.
$\varepsilon_0^{d(p)}$ is the $d(p)$ onsite energy of Mn $e_g$ (O-$p$), whereas $t^{pp}_{j,j'}$ and $
V_{i,j}^{\alpha}$ are hopping parameters that describe hybridization between neighboring O $p$ states and
between Mn $e_g$ states and O $p$ states, respectively.

In order to get a faithful estimate of the onsite energies and the hopping interactions, 
we have carried out NMTO-downfolding calculations\cite{andersen_nmto}. Starting from a full DFT calculation for 
cubic CaMnO$_3$, NMTO-downfolding arrives at a 5-orbital Hamiltonian by integrating out degrees of freedom 
involving Mn-$t_{2g}$, O-$p_{\pi}$ and those involving Ca. The downfolded 5-orbital Hamiltonian is built by defining energy-selected, effective Mn-$e_g$ and O-$p_{\sigma}$ 
orbitals which serve as Wannier-like orbitals spanning the downfolded five bands.
Linear muffin-tin calculations (LMTO)\cite{andersen_lmto} were performed in order to evaluate 
self-consistently the potential parameters needed for the NMTO procedure.

The second and third term in Eqn. (\ref{ham:mn}) describe the Coulomb interactions in the $3d$ shell of the $TM$ cation.
Treating the $t_{2g}$ electrons as classical spins ${\bm S}_i$, the interaction between them and the $e_g$ electrons can be 
accounted for by the Hund's coupling $J_H$, connecting the localized core spins ${\bm S_i}$ with the spins ${\bm s}_i=\sum_{\alpha,\sigma,\sigma'}\,
d^\dagger_{i\alpha\sigma}\, \hat{\bm \tau}_{\sigma\sigma'}\,d^{\phantom{\dagger}}_{i\alpha\sigma'}$ of the mobile electrons,
$\hat{\bm\tau}$ being the Pauli matrices, as given in the second term in Eqn. (\ref{ham:mn}). In most of our
calculations, we have assumed the antiferromagnetic G-type arrangement of $t_{2g}$ spins, though to study the influence of
spin arrangements, we have also considered the ferromagnetic alignment of $t_{2g}$ spins. The third term, $H_{e-e}$ describes 
the Coulomb interactions between $e_g$ electrons, as given below. We drop the site index for sake of simplicity.
\beq
H_{e-e} &=& \frac{U}{2}\,\sum_{\alpha,\sigma}\,n_{\alpha\sigma}n_{\alpha\bar{\sigma}}\,+\\
&&\,\frac{U'}{2}\,
\sum_{\alpha,\alpha',\sigma,\sigma'}\,(1-\delta_{\alpha\alpha'})\,n_{\alpha\sigma}\,n_{\alpha'\sigma'} + \nonumber\\
&&\frac{J}{2}\sum_{\alpha,\alpha',\sigma,\sigma'}\Bigl[ (1-\delta_{\alpha\alpha'})\,
d^\dagger_{\alpha\sigma}d^\dagger_{\alpha'\sigma'}d^{\phantom{\dagger}}_{\alpha\sigma'}
d^{\phantom{\dagger}}_{\alpha'\sigma} \, + \nonumber\\
&&\,(1-\delta_{\alpha\alpha'})(1-\delta_{\sigma\sigma'})\,
d^\dagger_{\alpha\sigma}d^\dagger_{\alpha\sigma'}d^{\phantom{\dagger}}_{\alpha'\sigma'}
d^{\phantom{\dagger}}_{\alpha'\sigma}\Bigr],  \nonumber
\eeq
$U,U',J$ are the intraband Coulomb, interband Coulomb and interband exchange couplings respectively\cite{kanamori}, defined 
within the $e_g$ manifold. The values of  $U,U',J$ were estimated as, $U=7.02~eV,\,U'=4.97~eV$
and $J=1.025~eV$, as relevant for CaMnO$_3$.\cite{elp}
Treating $H_{e-e}$ at the mean-field level, applying a standard Hartree-Fock linearization of two-body operators,
the problem reduces to a single-particle problem, which can be solved straightforwardly in reciprocal space.
Considering the fact that the $e_g$ manifold is empty, such a mean-field treatment is justified.
In order to take
into account the G-type antiferromagnetic configuration of $t_{2g}$ spins, we defined a unit cell hosting two $TM$ cations
and six O ions\cite{footnote2}. After transformation in reciprocal space, the Hamiltonian can be diagonalized for any given value of momentum
$k$; integration was then performed using a nested Clenshaw-Curtis quadrature rule\cite{clenshawcurtis}.

In order to investigate the ferroelectric instability, we considered an off-centering displacement $u$ of each Mn 
cation along the $+\hat{z}$ direction, {\it i.e.}, towards one of the apical oxygen, as considered 
in Ref. \onlinecite{ederer}. The corresponding hopping parameters were modified proportionally to $\sim
(d/d_0)^{-3.5}$, $d$ and $d_0$ being the distorted and undistorted Mn-O bond lengths\cite{harrison} respectively.\cite{footnote1}
The elastic energy loss associated with the distortion, $u$, was evaluated using total-energy DFT calculations,
as mentioned earlier. The plane wave calculations, carried out for this purpose, were performed using
projector augmented wave (PAW) potentials\cite{pseudo} and the wavefunctions
were expanded in the plane wave basis with a kinetic energy cut-off of 450 eV. Reciprocal space integration
was carried out with a k-space mesh of 6$\times$6$\times$6. 

\section{Results}\label{section:results}

\subsection{Volume effects on covalency energy, role of correlation and Hund's coupling}

Using the NMTO-downfolding technique, as explained in Section II, we evaluated the onsite energies and various 
hopping interactions, relevant for CaMnO$_3$ for two cubic crystal structures, one at equilibrium
volume\cite{wollan} with lattice constant $a_0=3.73$ \AA~ and another at increased volume, as under a negative isotropic
pressure, with  $2\%~$ increased lattice constant $a_0=3.805~$\AA. The obtained parameters are listed in Table \ref{tbl1}.
The hopping interactions 
between Mn and O sites is given by, $V_{ij}^\alpha=(-1)^{M_{ij}}\,t^{\alpha\beta}_{pd}$, where $M_{ij}=2$ if $j=i+\hat{e}/2$ and
$M_{ij}=1$ if $j=i-\hat{e}/2$; $\beta=x,y,z$ labels the $p$ orbital state that hybridize with Mn $e_g$ states. Within a cubic symmetry, the interaction between $TM$  $x^2-y^2$ and O $p_z$ is zero, and therefore not listed in the table.

\begin{table}[h!]
\vspace{-0.1cm}
\begin{tabular}{c|ccccccc}
%\hline
\hline
 & $\varepsilon_d$  & $\varepsilon_p$ & $t_{pd}^{1,z}$ & $t_{pd}^{1,x(y)}$ &$t_{pd}^{2,x(y)}$ &$t^{pp}$\\
 \hline
 eq & -2.86. & -5.16 & 1.95 &1.69 & 0.98 & 0.75 \\
 2\%& -3.10 & -5.03  & 1.76 & 1.53 & 0.88 &0.59\\

 \hline
% \hline
\end{tabular}
\caption{Onsite and hopping parameters at equilibrium and $2\%$ increased lattice constant, obtained from 
NMTO-downfolding calculations.}\label{tbl1}
\vspace{-0.2cm}
\end{table}

Naively, one would expect that the tendency to form strong covalent bonds would decrease in the expanded volume structure.
As is seen from the parameters listed in Table \ref{tbl1}, the hopping parameters are reduced by $\sim 10\%$
upon 2$\%$ increase in lattice constant. However, interestingly, the volume expansion also influences the level splitting
between $d$ and $p$ states, as is seen in Table \ref{tbl1}. Following Khomskii\cite{khomski1}, one can estimate the 
energy gain due to the off-centering of the $TM$ ion, as described below. When a Mn atom shifts towards an apical oxygen, 
reducing the corresponding Mn-O bond length, it forms a strong covalent bond with this particular oxygen at the expense 
of weakening the bonds with other oxygen ions. Within a linear approximation and neglecting the bonds between the Mn ion 
and the in-plane oxygens, whose bond lengths are only slightly increased in the considered off-center shift, the covalency
energy gain can be estimated as $\delta E_{cov} \approx -2\,(u\,t_{pd})^2/\Delta$, $\Delta$ being the charge-transfer energy,
given by $\vert\varepsilon_d-\varepsilon_p\vert$. An estimate of the changes in covalency energy gain induced by volume expansion 
may be obtained by considering the ratio $t_{pd}^2/\Delta$, rather than hopping parameters $t_{pd}$ only. As given in Table \ref{tbl1}, 
upon volume expansion $\Delta$ is reduced by $\sim 16\%$, giving rise to a compensating effect of the reduction
in hopping interaction.  This leads to only a reduction by less than $4\%~$ in covalency energy gain upon volume
expansion.

\begin{figure}[h]
\includegraphics[width=8.8cm]{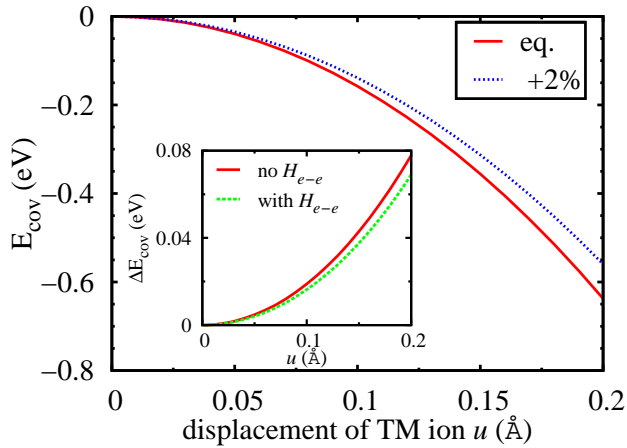}
\caption{Covalency energy in the 5-band model as a function of off-centering shift of Mn cation towards the apical
oxygen at equilibrium and $2\%$ increased lattice parameters, for $H_{e-e}$ set to zero. Energies are shifted in
such a way that $E_{cov}=0$ at $u=0$. The negative values indicate a gain in covalency energy upon off centric movement. 
The inset shows the difference $E_{cov}(2\%)-E_{cov}(eq.)$ with and without $H_{e-e}$ but neglecting Hund's coupling.}\label{fig1}
\end{figure}

This above expectation is confirmed by our calculations on the model Hamiltonian. We carried out two sets of
calculation, one excluding $H_{e-e}$ and another including $H_{e-e}$. In both set of calculations, we assumed $J_H=0$, therefore
Hund's coupling was not operative. The results are shown in Fig. \ref{fig1}. As discussed in the above, the volume expansion
was found to reduce the covalency energy gain due to off-centric movement marginally. Inclusion of correlation effects 
between $e_g$ electrons, keeping $J_H=0$, lead to a further suppression of the volume effect. This presumably is
rationalized by
the fact that inclusion of Coulomb interactions would tend to suppress hybridization interaction between oxygens and $TM$ states, 
thus also making the effect of volume variation on hybridization energy gain less effective.

\begin{figure}[h]
\includegraphics[width=8.8cm]{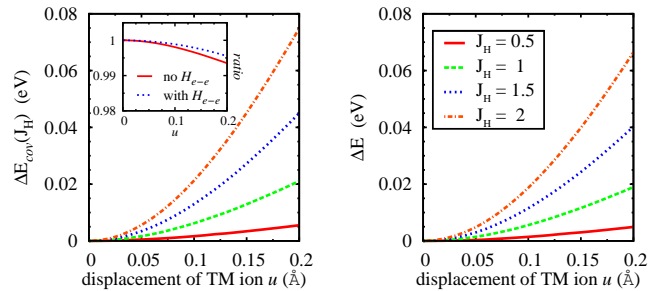}
\caption{Effect of Hund's coupling on the covalency energy at equilibrium (left) and expanded (right) volume as a function
of the displacement $u$ measured by the energy difference $\Delta E_{cov}$, neglecting other
Coulomb interactions. The inset in the left panel shows the
ratio $E_{cov}(u;J_H=2)/E_{cov}(u;J_H=0)$, measuring the relative energy loss with and without $H_{e-e}$.}\label{fig2}
\end{figure}

In the next step, we investigated the role of Hund's coupling between mobile electrons and the localized spins on Mn ions.
For this purpose, we evaluated the energy gain induced by the off-centering distortions at selected values of $J_H$, thus 
emphasizing its effect on the hybridization mechanism. We assumed G-type antiferromagnetic ordering for $t_{2g}$ spins,
which is the magnetic ordering observed for CaMnO$_3$.
A realistic value for the Hund's coupling in manganites 
should be\cite{elp,dagotto.rev}  between $1$ and $2~eV$. We have varied the $J_H$ value from $0.5$ to $2~eV$. 
As proposed by Khomskii\cite{khomski1}, we find that the presence of local spins on the $TM$
ions and their coupling to the mobile electrons through Hund's interaction reduces the energy gain due to the
off centering movement. As discussed in Ref.\onlinecite{khomski1}, the strong covalent bond between $TM$ $e_g$ states 
and neighboring oxygen $p$ states implies the formation of a singlet state. This is disfavoured by the strong Hund's coupling with 
the localized $t_{2g}$ electrons on $TM$ ion; this interaction can be seen as a local Zeeman field due to core spins and effectively
acts as a ``pair-breaker'' on the singlet state\cite{khomski1}.
This is demonstrated in the left panel of Fig. \ref{fig2}, where we plot the energy 
difference $\Delta E_{cov}=[E_{cov}(u;J_H)-E_{cov}(u;J_H=0)]$ as a function of the off-centering distortion $u$. 
We find this energy difference to be positive, indicating that introduction of $J_H$ reduces the energy gain
due to the off-centering movement. Increasing $J_H$ reduces more and more the covalency energy,
making $\Delta E_{cov}$ more positive. Inclusion of electron-electron correlation within $e_g$ manifold,
through inclusion of $H_{e-e}$ term, reduces the overall covalency energy due to off centering movement, as discussed
in the context of  Fig. \ref{fig1}. However, the relative energy loss in covalency energy 
due to the coupling with the underlying local spins, expressed as $E_{cov}(u;J_H=2)/E_{cov}(u;J_H=0)$,
remains similar between calculations excluding and including $H_{e-e}$ term, as shown in the inset of left panel of 
Fig. \ref{fig2}. The calculations repeated for the 2$\%$ expanded lattice constant, exhibit the same trend, as shown in
the right panel of Fig. \ref{fig2}. This leads us to conclude that the effectiveness or role of Hund's coupling
to be nearly same for the equilibrium and the expanded volume. 

\begin{figure}[h]
\includegraphics[width=8.8cm]{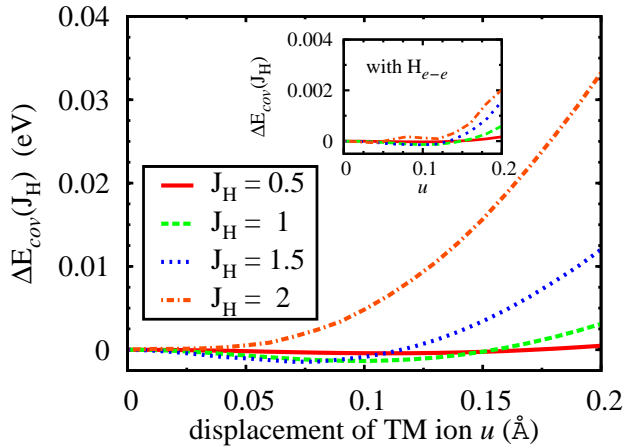}
\caption{Effect of Hund's coupling on the covalent energy gain at equilibrium volume as a function of the displacement $u$
for a ferromagnetic ordering of local $t_{2g}$ spins. The inset shows the same including $H_{e-e}$.}\label{fig3}
\end{figure}

The effectiveness of Hund's coupling, on the other hand, is found to depend strongly on the magnetic arrangements
of the underlying $t_{2g}$ spins. The above discussed set of calculations have been repeated for the ferromagnetic
alignment of the $t_{2g}$ spins. The results are summarized in Fig. \ref{fig3} for the equilibrium volume. 
Comparing the values of  $\Delta E_{cov}$, we find them to be much smaller than those calculated
for the G-type antiferromagnetic ordering of $t_{2g}$ spins and shown in Fig. \ref{fig2},
suggesting that Hund's coupling is much less effective.
$\Delta E_{cov}$ is even found to attain small, negative values for $J_H$ between $1 - 1.5 eV$ for moderate values
of off centering movement. Inclusion of $H_{e-e}$, makes the effect of $J_H$ even milder, with much smaller values
of $\Delta E_{cov}$, as shown in the inset of Fig. \ref{fig3}. This finding is in accordance with the predicted jump
in polarization of epitaxially strained SrMnO$_3$, when the system undergoes an antiferromagnetic-ferromagnetic phase
transition\cite{leerabe}.

\subsection{Estimation of elastic stiffness. Volume effect on total energy}

\begin{figure}[h]
\includegraphics[width=8cm]{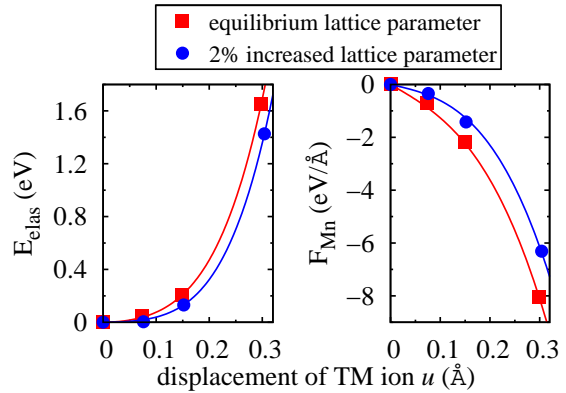}
\caption{The elastic energy (left panel) and force acting on Mn ion (right panel) plotted as a function of off centric displacement,
obtained from DFT calculations.}\label{fig4}
\end{figure}

The analysis, so far, has been restricted to the discussion of covalency energy gain due to off centering movement of $TM$ cation.
In order to arrive to a complete picture of polar instability, one needs to take into account the restoring force, which 
is given by the elastic stiffness associated to the displacement of $TM$ cation.
The elastic energy, that emerges from the short-range repulsive interactions between ions, for small displacements, 
is expected to go as $\sim k\,u^2/2$, $k$ being the stiffness constant. We estimated the stiffness constant from total
energy calculations carried out on plane wave basis. One may identify the covalency energy with the band energy contribution 
to the total energy, and the elastic energy as the remainder, {\it i.e.} $E_{tot}=E_{cov}+E_{elas}$.
The relevant stiffness constant can then be extracted via a fitting procedure from the evolution of $E_{elas}$ as a function of 
the shift $u$. This is shown in left panel of Fig. \ref{fig4} for both the equilibrium and expanded volume. The fitting 
function was chosen as $f(u)=ku^2/2+cu^4$. As shown in Table \ref{tbl2}, while the coefficient of the quartic term does not change 
in the expanded volume system with respect to the equilibrium case, the stiffness coefficient $k$ is reduced by a factor of 0.4 
in the expanded volume, suggesting a strong reduction of the repulsive short-range interactions.

\begin{table}[h!]
\vspace{-0.1cm}
\begin{tabular}{c|cc|cc}
\hline
%\hline
 & eq.   & $2\%$  & eq.  & $2\%$ \\
 & \hspace{1.5cm}($E_{elas}$)   &                & \hspace{1.5cm}($F_{TM}$)&                \\
 \hline
 $k\, (eV/$\AA$^2)$ & 12.365 & 4.961 & 10.846 & 4.291 \\
 $c\, (eV/$\AA$^4)$ & 140 & 140  & 45 & 45  \\
% \hline
 \hline
\end{tabular}
\caption{Stiffness constants as evaluated from estimates from elastic energies at equilibrium and $2\%$ increased lattice parameters
(shown in first and second columns respectively) and as evaluated from force acting on Mn cation (shown in 
third and fourth columns respectively).}\label{tbl2}
\vspace{-0.2cm}
\end{table}

In order to check the goodness of our estimate, we also adopted an alternative procedure of extracting the stiffness constant 
from the force acting on the $TM$ ion. Right panel of Fig. \ref{fig4} shows the evolution of the force acting on Mn ion, $F_{Mn}$
as a function of the off centered displacement. The calculated data points were fitted with the function $-f^\prime(u)=-ku-4cu^3$.
As shown in Table \ref{tbl2}, the estimated $k$ coefficient
shows same order of magnitude and same trend under volume expansion, whereas the coefficient of the quartic term is more
than three times smaller when evaluated from the total force. This finding, however, is not surprising if we notice that, in
this second approach, we are neglecting forces acting on apical oxygens when we artificially shift the Mn
cation. We, however, do not expect the physical picture to be change much between the two procedures, since in both 
the approaches, while the coefficient $k$ was found to show a significant reduction under volume expansion, the coefficient of 
the quartic term, $c$, was found to remain unchanged by volume expansion.

\begin{figure}[h]
\includegraphics[width=8.8cm]{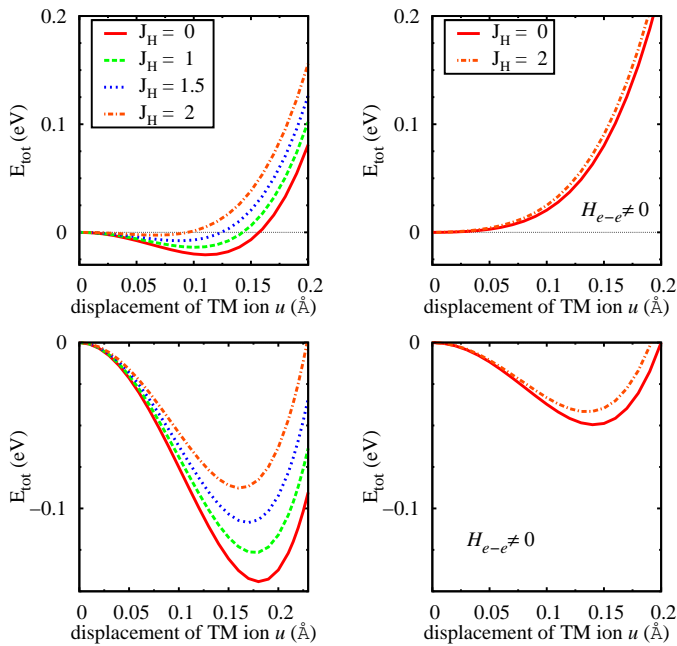}
\caption{ Total energy calculated from the model Hamiltonian as a function of the off-centering shift of $TM$
ion at equilibrium (top panels) and $2\%$ increased lattice parameter (bottom panels) for different choice of
Hund's coupling. Correlations within $e_g$ electrons have been set to zero in left top and bottom panels, while they are
included at the mean-field level in right top and bottom panels.
}\label{fig5}
\end{figure}

From the knowledge of the stiffness constant and covalency energies, in the next step, we investigated 
the ferroelectric instability in cubic perovskite structure of CaMnO$_3$ which arises due to the competition 
between the covalency energy and the elastic energy in the model Hamiltonian framework. With the estimated parameters
discussed in previous sections, we find a weak ferroelectric instability for the system at equilibrium volume, which
gets weakened more and more as we allow for a finite Hund's coupling between mobile electrons participating in the 
hybridization mechanism and the local Mn spins, finally nearly disappearing for $J_H=2~ eV~$, as shown in left 
top
panel of Fig. \ref{fig5}. Upon increasing volume, however, a strong instability appears, which survives even after 
inclusion of Hund's coupling, though gets shifted to lower values of off centric displacement,
as shown in left bottom panel of  Fig. \ref{fig5}. In the above calculations, $H_{e-e}$ was set to zero.
Inclusion of $H_{e-e}$,
as discussed in the above, weakens the covalency effect and the ferroelectric instability completely vanishes
for the equilibrium volume, as shown in the right panels of Fig. \ref{fig5}. The ferroelectric instability, however,
remains in the expanded volume, even for largest Hund's coupling strength of $2\,eV$.

\section{Conclusions}

To conclude, we have introduced a minimal 5-band $d-p$ model with Coulomb interactions on the $TM$ ion
to describe the possible hybridization mechanism for ferroelectricity in magnetic Mn-based perovskites.
As a case study, we have considered the cubic perovskite structure of CaMnO$_3$ and have considered
two different volumes of the structure, the equilibrium volume and 2$\%$ expanded volume. 
Our study shows that the ferroelectric instability crucially depends on the balance between two main energy contributions, one 
coming from the tendency of empty $d$ states to form covalent bonds with surrounding O states and the other from the
short-range forces between ions, resulting in an elastic energy associated to the $TM$ off-centering shift. 
We found that volume expansion has marginal effect on the covalency energy gain, since the reduction in hopping
interaction gets compensated by the simultaneous gain due to change in charge transfer energy. The coupling of 
the mobile electrons with localized spins on magnetic ions through Hund's interaction destabilizes the gain
in covalency energy due to off centric movement, as has been suggested by Khomskii\cite{khomski1}. Interestingly,
we find the influence of the Hund's coupling to crucially depend on the spin arrangements of the localized $t_{2g}$ 
spins, suggesting a rather strong spin-phonon coupling, as predicted in epitaxially strained SrMnO$_3$\cite{leerabe}.
Since the hybridization driven mechanism turned out to be weakly volume dependent, the effectiveness of 
Hund's coupling in reducing the gain in covalency energy, which is a local effect, 
is found to be weakly affected by volume expansion. 
On the contrary, we found that the stiffness of the Mn off-centering
displacement strongly depends on the volume, being more than halved for a $2\%$ increase in lattice parameter. The strong reduction of 
short-range forces, combined with an almost unaffected tendency to covalency leads to stronger ferroelectric instability
in expanded volume. The correlation between the $e_g$ electrons has the role of suppressing the covalency effect, which washes 
out the weak ferroelectric instability found for the equilibrium volume, but retains the much more pronounced ferroelectric instability
for the expanded volume.

Our results highlight the crucial role played by the balance of hopping integrals and charge-transfer energy in determining 
the effectiveness of the covalency. For instance, much weaker ferroelectric instability predicted for the iso-electronic $d^3$ LaCrO$_3$
can be ascribed to a larger charge-transfer energy $\Delta$, as suggested in Ref. \onlinecite{ederer}. On the other hand, one
can speculate that the tendency to form covalent bonds is reduced for systems with partially filled $t_{2g}$ states, e.g.
with $d^1$ or $d^2$, even if Hund's coupling is expected to play a less effective role compared to $d^3$ configuration; in these systems, 
the empty $t_{2g}$ electrons would establish $\pi$ bonds with neighboring oxygens, implying smaller
hopping integrals, whereas the charge-transfer energy is generally expected to be larger than that for $d^3$ systems. The critical 
balance between different energy contributions needs to be estimated as in present case. It would, therefore, be interesting
to repeat the above analysis for other $d^n$ perovskite with $n < 3$.

Finally, it will be worthwhile to comment on the following. The ferroelectric 
instability in perovskite oxides is usually explained\cite{cochran} in terms of a close 
competition between short-range (SR) forces, which tend to favor the non-polar cubic structure, and
the long-range (LR) interactions, which tend to favor the ferroelectric phase. 
In particular, in Ref. \onlinecite{ghosez} such SR and LR 
contributions  have been discussed in the context of CaMnO$_3$. It may
therefore be interesting to connect the various energy contributions
discussed in the present work to SR and LR forces. Since the Hund's coupling
as well as the restoring elastic stiffness tend to destabilize the
ferroelectricity, they may be linked to SR forces. 
As detailed in Ref. \onlinecite{ghosez}, the LR
contribution has been shown to be directly related to anomalously large Born 
effective charges (see also Ref. \onlinecite{ghosez1996}).
 We note that the presence of anomalously large charges
originates from a change in the hybridization between occupied and unoccupied
electronic states\cite{ghosez1998}.
In the context of CaMnO$_3$, this translates to the change in hybridization 
or covalency between empty Mn e$_g$ states and occupied O-$p$ states, which 
is in turn determined by both the hopping interaction, $t_{pd}$, and the charge transfer 
energy, $\Delta$, both affecting the LR forces.

{\it Acknowledgements} --
We thank D. Khomskii for reading the manuscript and for his helpful insights.
This work has been supported by the European Community's Seventh Framework Programme FP7/2007-2013 under grant agreement
No. 203523-BISMUTH. The project started during the AQUIFER research
program held at CNR-SPIN L'Aquila, Italy in September-October 2010 and sponsored by the NSF International Center for
Materials Research at UCSB, USA. TSD would like to thank Advanced Materials Research Unit (AMRU)
for financial assistance.


\begin{thebibliography}{200}
\bibitem{footnote} In the perfect cubic structure $t_{2g}$ electrons hybridize mainly with O $p-\pi$ orbitals, which do not hybridize with the
$e_g$; this independence of bonding mechanisms further supports our assumption of a 5-band model with localized $t_{2g}$ spins.
\bibitem{footnote2} In order to enforce the G-type antiferromagnetic configuration, we take advantage of the lattice being
bipartite within cubic structure and assume staggered periodic boundary
conditions, in such a way that a site belonging to sublattice A is connected only to sites belonging to
sublattice B, both within the defined unit cell and through the boundaries.
\bibitem{footnote1} We considered only changes in $\sigma$ bonding, neglecting possible small contributions from $\pi$ bonding energies.
Considering $\pi$ bonding would imply extending the present 5-band model to an 11-band model taking into account all three $p$ orbitals on
each oxygen. We have checked the validity of our 5-band model results, in terms of comparison with calculations of 11-band model. We find the 
conclusions to remain unchanged with main changes to covalency energy coming from hybridization with apical oxygens, given by $\sigma$ bonding.
\end{thebibliography}
\end{document}